# Determining the three-dimensional atomic structure of a metallic glass

Yao Yang[1*], Jihan Zhou[1*], Fan Zhu[1*], Yakun Yuan[1*], Dillan J. Chang[1], Dennis S. Kim[1], Minh Pham[2], Arjun Rana[1], Xuezeng Tian[1], Yonggang Yao[3], Stanley J. Osher[2], Andreas K. Schmid[4], Liangbing Hu[3], Peter Ercius[4] & Jianwei Miao[1]

[1]*Department of Physics & Astronomy, STROBE NSF Science & Technology Center and California NanoSystems Institute, University of California, Los Angeles, CA 90095, USA.*
[2]*Department of Mathematics, University of California, Los Angeles, CA 90095, USA.*
[3]*Department of Materials Science and Engineering, University of Maryland, College Park, Maryland, 20742, USA.* [4]*National Center for Electron Microscopy, Molecular Foundry, Lawrence Berkeley National Laboratory, Berkeley, CA 94720, USA.*

*These authors contributed equally to this work.

**Amorphous solids such as glass are ubiquitous in our daily life and have found broad applications ranging from window glass and solar cells to telecommunications and transformer cores[1,2]. However, due to the lack of long-range order, the three-dimensional (3D) atomic structure of amorphous solids have thus far defied any direct experimental determination without model fitting[3-13]. Here, using a multi-component metallic glass as a proof-of-principle, we advance atomic electron tomography to determine the 3D atomic positions in an amorphous solid for the first time. We quantitatively characterize the short-range order (SRO) and medium-range order (MRO) of the 3D atomic arrangement. We find that although the 3D atomic packing of the SRO is geometrically disordered, some SRO connect with each other to form crystal-like networks and give rise to MRO. We identify four crystal-like MRO networks − face-centred cubic, hexagonal close-packed, body-centered**

**cubic and simple cubic – coexisting in the sample, which show translational but no orientational order. These observations confirm that the 3D atomic structure in some parts of the sample is consistent with the efficient cluster packing model[8,10-12,20]. Looking forward, we anticipate this experiment will open the door to determining the 3D atomic coordinates of various amorphous solids, whose impact on non-crystalline solids may be comparable to the first 3D crystal structure solved by x-ray crystallography over a century ago[14].**

Since the first discovery in 1960[15], metallic glasses have been actively studied for fundamental interest and practical applications[7-12,16-20]. However, due to their disordered structure, the 3D atomic arrangement of metallic glasses cannot be determined by crystallography[21]. Over the years, a number of experimental and computational methods have been used to study the metallic glass structure, such as x-ray/neutron diffraction[22,23], x-ray absorption fine structure[9], high-resolution transmission electron microscopy[24], fluctuation electron microscopy[25], angstrom- and nano-beam electron diffraction[13,26,27], nuclear magnetic resonance[28], density functional theory[29], molecular dynamics simulations[30-33] and reverse Monte Carlo modelling[9,25]. Despite all these developments, however, there was no experimental method available to directly determine all the 3D atomic positions in metallic glass samples. One experimental method that can potentially solve this long-standing problem is atomic electron tomography (AET)[34,35]. AET combines high-resolution tomographic tilt series with advanced iterative algorithms to resolve the 3D atomic structure of materials without assuming crystallinity, which has been applied to image grain boundaries, anti-phase boundaries, stacking faults, dislocations, point defects, chemical order/disorder, atomic-scale ripples, bond distortion and strain tensors with unprecedented 3D detail[36-41]. More recently, 4D (3D + time) AET has been developed to observe crystal nucleation at atomic resolution, showing that early stage nucleation results are inconsistent with classical nucleation theory[42]. Here, we use a multi-component metallic glass as a model and advance AET to determine its 3D atomic positions with a precision of 21 picometers.

**Determining the 3D atomic positions in a multi-component metallic glass**

The samples were synthesized by a carbothermal shock technique with a high cooling rate (Supplementary Fig. 1a, Supplementary video 1 and Methods), which created high entropy alloy nanoparticles with multi-metal components[43]. The energy-dispersive X-ray spectroscopy data show the nanoparticles are composed of eight elements: Co, Ni, Ru, Rh, Pd, Ag, Ir and Pt (Supplementary Fig. 1b-k). Tomographic tilt series were acquired from seven nanoparticles using a scanning transmission electron microscope with an annular dark-field detector (Supplementary Table 1). While most of the nanoparticles are crystalline or polycrystalline, particles 1 and 2 have disordered structure (Supplementary Fig. 2). In this study, we focus on the more disordered nanoparticle (particle 1), from which a tilt series of 55 images were measured (Fig. 1a and Supplementary Fig. 3a). Although some crystalline features are present in several images, the diffraction patterns calculated from the images show the amorphous halo. From the average diffraction pattern (Fig. 1b), we derived the radial distribution function (RDF) (Supplementary Fig. 3c), exhibiting the amorphous structure of the nanoparticle. After pre-processing and image denoising, the tilt series was reconstructed and the 3D atomic positions were traced and classified (Fig. 1c, d, Supplementary Video 2 and Methods). Since the image contrast in the 3D reconstruction depends on the atomic number[40-42], presently AET is only sensitive enough to classify the eight elements into three different types: Co and Ni as type 1, Ru, Rh, Pd and Ag as type 2, and Ir and Pt as type 3. After atom classification, we obtained the 3D atomic model of the nanoparticle, consisting of 8322, 6896 and 3138 atoms for type 1, 2 and 3, respectively. To verify the reconstruction, atom tracing and classification procedure, we calculated 55 images from the experimental atomic model using multislice simulations (Methods). Supplementary Fig. 4c and d shows the consistency between the experimental and calculated images. We then applied the reconstruction, atom tracing and classification procedure to obtain a new 3D atomic model from the 55 multislice images. By comparing the two models, we estimated that 97.37% of atoms were correctly identified with a 3D precision of 21 pm (Methods and Supplementary Fig. 4e).

Figure 1e and Supplementary video 3 show the experimental 3D atomic model of the nanoparticle with type 1, 2 and 3 atoms in green, blue and red, respectively. To quantitatively characterize the atomic structure, we employed the local bond orientational order (BOO) parameters to distinguish between the disordered, face-centred cubic (fcc), hexagonal close-packed (hcp) and body-centered cubic (bcc) structures (Methods). Figure 1f shows the local BOO parameters of all the atoms in the nanoparticle, indicating the majority of atoms severely deviate from the fcc, hcp and bcc crystal structures. For a comparison, the local BOO parameters of the seven nanoparticles are shown in Supplementary Fig. 2h-n. To separate crystal nuclei from the amorphous structure, we used the normalized BOO parameter to identify the crystal nuclei (Methods). By choosing the criterion of the normalized BOO parameter $\geq 0.5$ (Supplementary Fig. 2o), we identified 15.46% of the total atoms forming crystal nuclei in the nanoparticle (Supplementary Fig5a), which contribute to the crystalline features observed in several images (Supplementary Fig. 3a). In the following sections, we focus on the analysis of the atoms with the normalized BOO parameter $< 0.5$.

Figure 1g shows the RDF of the amorphous structure of the 3D atomic model (Methods), where the weak second-peak splitting is consistent with previous observation in high entropy bulk metallic glasses[44]. The ratios of the second, third, fourth and fifth to the first peak position are 1.74, 1.99, 2.64 and 3.51, respectively, which agree with those of metallic glasses[45,46]. The partial pair distribution functions (PDFs) between type 1, 2 and 3 atoms are shown in Fig. 1h. By fitting a Gaussian to the first peaks in the partial PDFs, we determined the type 11, 12, 13, 22, 23 and 33 bond lengths to be 2.59, 2.71, 2.78, 2.72, 2.75 and 2.9 Å, respectively. In particular, the partial PDF for the type 33 atoms (the yellow curve) exhibits a unique feature with the second peak higher than the first peak, indicating that the majority of type 3 atoms are distributed beyond the SRO.

**The short-range order**

To determine the SRO in the metallic glass sample, we used the Voronoi tessellation to characterize the local atomic arrangement[6]. This method identifies the nearest neighbour atoms around each central atom to form a Voronoi polyhedron, which is designated by a Voronoi index <$n_3$, $n_4$, $n_5$, $n_6$> with $n_i$ denoting the number of $i$-edged faces. Figure 2a shows the ten most abundant Voronoi polyhedra in the nanoparticle with a fraction ranging from 5.02% to 1.72%, most of which are geometrically disordered and commonly observed in model metallic glasses[11] such as <0,4,4,3>, <0,3,6,3>, <0,4,4,2> and <0,3,6,2> (Fig. 2b). The small fractions of the Voronoi polyhedra suggest that the sample is not a well relaxed metallic glass due to its poor glass forming ability[18]. Figure 2c shows the local symmetry distribution of all the faces of the Voronoi polyhedra. The 3-, 4-, 5- and 6-edged faces account for 3.27%, 29.14%, 43.91% and 23.67%, respectively, revealing that 5-edged faces are most abundant in the SRO. But only 7.03% of all the Voronoi polyhedra are distorted icosahedra, including Voronoi indices <0,0,12,0>,

<0,1,10,2>, <0,2,8,2> and <0,2,8,1>. This observation indicates that most 5-edged faces do not form distorted icosahedra in this metallic glass nanoparticle. From the Voronoi tessellation, we also calculated the distribution of the coordination number (CN) (Fig. 2d and Methods), where the average CNs for types 1, 2 and 3 atoms are 11.97, 12.02 and 12.41, respectively. Based on the partial CNs (Supplementary Fig. 5b), we quantified the chemical SRO using the Warren–Cowley parameters (Methods), indicating that the type 11 and 23 bonds are favoured, but the type 12 and 33 bonds are unfavoured. These results are consistent with the observations of the shortening of the type 11 and 23 bonds and the lengthening of the type 12 and 33 bonds (Methods).

**The medium-range order**

From the partial PDF of type 33 atoms (Fig. 1h, the yellow curve), we observed that the highest peak is located at 4.77 Å and 1.49 times higher than the nearest neighbour peak. This result indicates that the majority of type 3 atoms are distributed in the second coordination shell, which is between the first (3.86 Å) and the second minimum (6.08 Å) of the RDF curve (Fig. 1g). According to the efficient cluster packing model[8,10-12,20], solute atoms are surrounded by solvent atoms to form solute-centred clusters. These solute-centred clusters act as the basic building blocks and are densely packed in 3D space to constitute the MRO of metallic glasses. To quantitatively test this model with experimental data, we examined the distribution of the type 3 atoms in the second coordination shell and identified 85.47% of type 3 atoms as solute centres (Supplementary Fig. 5c and Methods). These solute centres are surrounded mainly by type 1 and 2 solvent atoms to form atomic clusters. Supplementary Fig. 5d shows the ten most abundant Voronoi polyhedra of the solute-centred clusters. The solute-centred clusters connect with each other by sharing one (a vertex), two (an edge) and three atoms (a face) as well as protrude into each other by sharing four and five atoms (Fig. 3a-e). Figure 3f shows the statistical distribution of the number of the solute-centred cluster pairs, which share from one to five atoms.

To locate the MRO in the metallic glass nanoparticle, we implemented a breadth-first search algorithm to look for the fcc-, hcp-, bcc-, simple cubic (sc-) and icosahedral-like structures of the solute centres (Methods). This algorithm globally searches for the MRO with a maximum number of solute centres. Each MRO is defined to have five or more solute centres with each solute centre falling within a 0.75 Å radius to the fcc, hcp, bcc, sc lattice or icosahedral vertices. We found that four types of MRO (fcc-, hcp-, bcc- and sc-like) coexist in the sample (Methods). Although we did not observe icosahedral-like MRO in this sample, our work does not rule out its existence in other metallic glasses[9]. Figure 3g shows the histogram of the four types of MRO as a function of the size (i.e. the number of solute centres), where the inset illustrates the fraction of the solute centre atoms in the four types of MRO. Figure 3h and Supplementary Video 4 show the 3D distribution of the MRO with each having eight solute centres or more. To verify our analysis, we also searched for the MRO with a 1 Å and 0.5 Å radius cut-off and observed the coexistence of the four types of MRO with different cut-off radii (Supplementary Figs. 6 and 7).

Next, we quantitatively characterized the MRO with a 0.75 Å radius cut-off. Figure 4a and b shows the length and volume distribution of the MRO in the metallic glass nanoparticle. The average length and volume of the fcc-, hcp-, bcc- and sc-like MRO were measured to be 2.27 ± 0.50, 2.40 ± 0.42, 2.07 ± 0.38, 2.11 ± 0.48 nm, 1.80 ± 0.64, 1.96 ± 0.53, 1.63 ± 0.46 and 1.96 ± 0.74 nm$^3$, respectively. Figure 5a, c, e and g shows four representative fcc-, hcp-, bcc- and sc-like MRO, where the solute-centred clusters exhibit only translational but no orientational order. To better visualize these MRO, the solute centres are orientated along the fcc, hcp, bcc and sc zone axes (Fig. 5b, d, f and h), showing that the 3D shapes of the MRO are anisotropic and the networks are distorted. We calculated the partial PDFs of all the fcc-, hcp-, bcc- and sc-like solute centres in the metallic glass nanoparticle and their corresponding maximum peak positions are at 4.62, 4.77, 4.82 and 3.88 Å, respectively (Fig. 4c). These peak positions represent the average nearest neighbour distances of the solute centres in the four crystal-like MRO and the broadened peaks signify the severe deviation of the MRO from perfect crystal lattices. Compared with the other three partial PDFs, the partial PDF of the sc-like MRO has two peaks and the ratio of the second to the first peak position is about $\sqrt{2}$ (Fig. 4c, the purple curve), which corresponds to the ratio of the diagonal to the side length of a square. The shorter nearest neighbour distance of the sc-like MRO than the other three crystal-like networks and the appearance of the two peaks in the partial PDF indicate that the sc-like solute-centred clusters are more closely connected with their neighbours. Figure 4d shows the distribution of sharing one, two, three, four and five atoms between neighbouring solute-centred clusters for the four types of the MRO, confirming that the solute-centred clusters in sc-like MRO tend to share more atoms with their neighbours than those in other types of MRO.

Our quantitative analysis of the SRO and MRO in a multi-component metallic glass provides direct experimental evidence to support the general framework of the efficient cluster packing model[8,10-12,20], that is, solute-centred clusters are densely packed in some parts of the sample to give rise to the MRO. We observed the chemical SRO, the bond shortening and lengthening, and the coexistence of fcc-, hcp-, bcc- and sc-like MRO in the multi-component metallic glass. By quantifying their length, volume and 3D structure, we found that the MRO not only has a large variation in length and volume, but also significantly deviates from perfect crystal lattices (Fig. 4c). As the size of MRO is comparable to that of shear transformation zones in metallic glasses[11,20,47,48], we expect that AET could also be applied to determine the 3D atomic structure of shear transformation zones and link the structure and properties of metallic glasses[11].

**Outlook**

Over the last century, crystallography has been broadly applied to determine the 3D atomic structure of crystalline samples[21]. The quantitative 3D structural information has been fundamental to the development of many scientific fields. However, for amorphous solids, their 3D structure has been primarily inferred from experimental data, where either the average statistical structural information can be obtained or model fitting is required to analyse the local atomic order[8-13]. These qualitative approaches have hindered our fundamental understanding of the 3D structure of amorphous solids and related phenomena such as the crystal-amorphous phase transition and the glass transition[11,49,50]. Here, we demonstrate the ability to directly determine the 3D atomic structure of an amorphous solid using AET, which enables us to quantitatively analyse the SRO and MRO at the single-atom level. Although we focus on a metallic glass nanoparticle in this study, this method is generally applicable to different sample geometries such as thin films and extended objects (Supplementary Figs. 8 and 9, Methods). Thus, we expect that this work could open a new era in determining the 3D structure of a wide range of amorphous solids.

**Acknowledgements** We thank J. Ding for stimulating discussions, H. B. Yu for providing a molecular dynamics simulated atomic model of the $Cu_{65}Zr_{35}$ metallic glass, and D. J. Kline and M. R. Zachariah for assistance with the temperature measurements. This work was primarily supported by STROBE: A National Science Foundation Science & Technology Center under Grant No. DMR 1548924. This work was also supported by the U.S. Department of Energy, Office of Science, Basic Energy Sciences, Division of Materials Sciences and Engineering under Award No. DE-SC0010378 (3D image reconstruction, atom tracing and classification, and multislice simulations) and the NSF DMREF program under Award No. DMR-1437263. J.M. acknowledges partial support from an Army Research Office MURI grant on Ab-Initio Solid-State Quantum Materials: Design, Production and Characterization at the Atomic Scale. The ADF-STEM imaging with TEAM 0.5 was performed at the Molecular Foundry, which is supported by the Office of Science, Office of Basic Energy Sciences of the U.S. DOE under Contract No. DE-AC02—05CH11231.

**Figures and Figure legends**

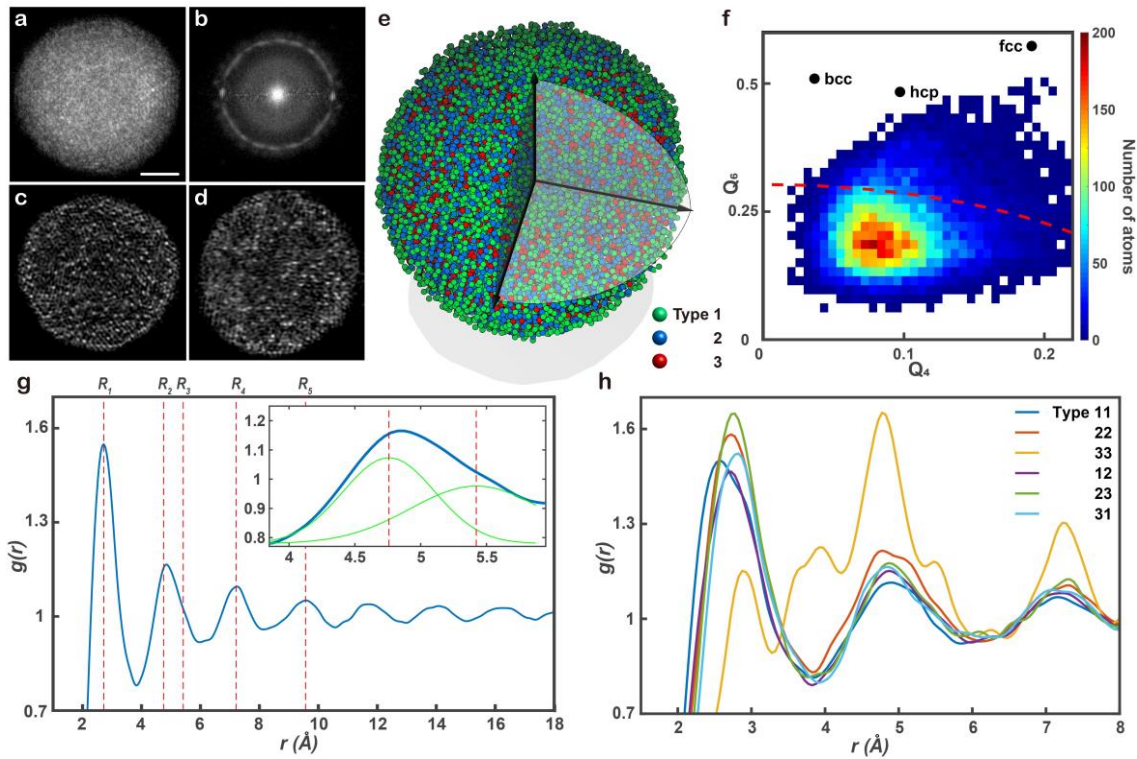

**Figure 1 | Determining the 3D atomic structure of a multi-component metallic glass with AET**. **a,** A representative experimental image, where some crystalline features are visible. **b**, Average diffraction pattern obtained from 55 experimental images (Supplementary Fig. 3a), showing the amorphous halo. **c**, **d**, Two 2.4-Å-thick slices of the 3D reconstruction in the *xy*- and *yz*-plane, respectively, where the majority of type 3 atoms (bright dots) are distributed in the second coordination shell. **e**, Experimental 3D atomic model of the metallic glass nanoparticle. **f**, The local BOO parameters of all the atoms in the nanoparticle. Based on the criterion of the normalized BOO parameter < 0.5 (the dashed red curve), 84.54% of the total atoms are disordered. **g**, The RDF of the

disordered atoms with the first, second, third, fourth and fifth peak positions ($R_1$, $R_2$, $R_3$, $R_4$ and $R_5$) at 2.73, 4.76, 5.42, 7.22 and 9.57 Å, respectively, which are in good agreement with those in the RDF directly derived from the average diffraction pattern (Supplementary Fig. 3c). The inset shows the second-peak splitting with a Gaussian fit. **h**, The partial PDFs between type 1, 2 and 3 atoms, consisting of 6 pairs - type 11, 12, 13, 22, 23 and 33 atoms. The partial PDF for the type 33 atoms (the yellow curve) shows a unique feature with a higher second peak than the first peak.

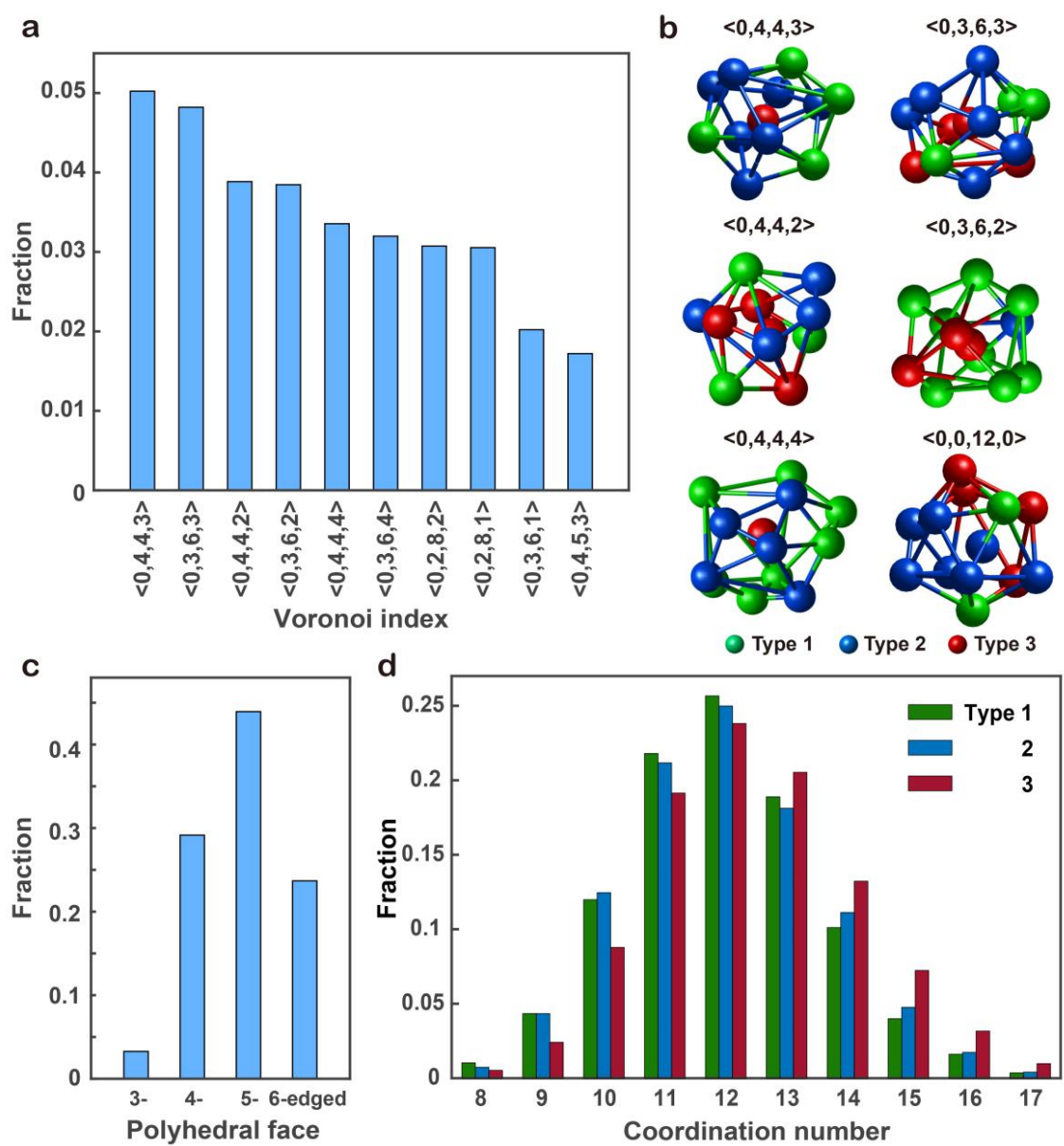

**Figure 2 | The short-range order of the metallic glass nanoparticle**. **a**, Ten most abundant Voronoi polyhedra in the nanoparticle. **b**, Six representative Voronoi polyhedra, where <0,4,4,3>, <0,3,6,3>, <0,4,4,2> and <0,3,6,2> are the four highest fraction Voronoi indices, <0,4,4,4> shows a severely distorted polyhedron, and <0,0,12,0> represents an icosahedron. **c**, The 3-, 4-, 5- and 6-edged face distribution of all the Voronoi polyhedra, where the 5-edged faces are the most abundant (43.91%). **d**, The coordination number (CN) distribution for type 1, 2 and 3 atoms.

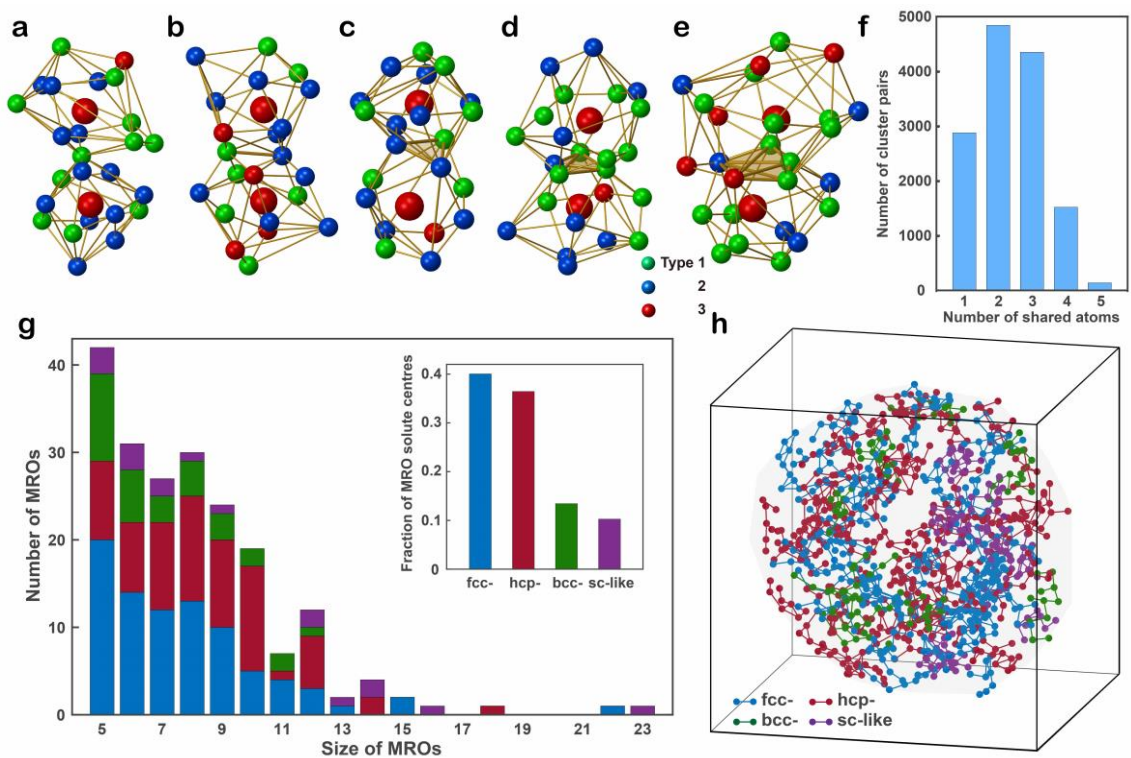

**Figure 3 | The connectivity and distribution of the MRO in the metallic glass nanoparticle**. **a-e**, Representative pairs of the solute-centred clusters that are connected with each other by sharing one, two, three, four and five atoms, respectively, where the central atom of each cluster is indicated by a large red sphere. **f**, Statistical distribution of the number of the solute-centred cluster pairs, which share from one to five atoms. **g**, Histogram of the four types of MRO – fcc- (in blue), hcp- (in red), bcc- (in green) and sc-

like (in purple) − as a function of the size (i.e. the number of solute centres). The total number of fcc-, hcp-, bcc- and sc-like MRO is 85, 71, 31 and 17, respectively. The inset shows the fraction of the solute centre atoms in the four types of MRO. **h**, Distribution of the four types of the MRO with eight solute centre atoms or more, where the centre region lacks of large MRO.

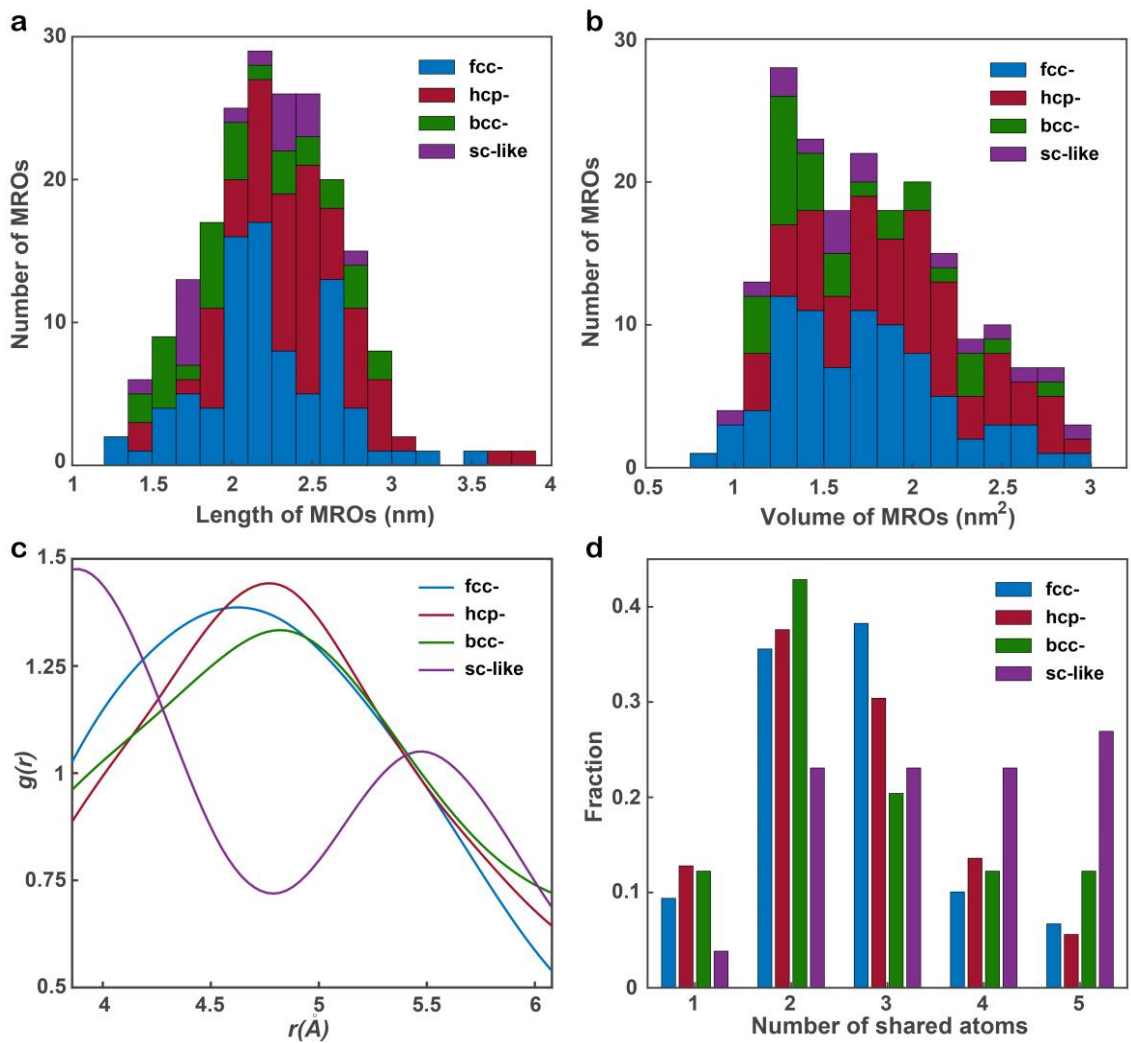

**Figure 4 | Quantitative characterization of the MRO.** The length (**a**) and volume (**b**) distribution of the four types of the MRO in the metallic glass nanoparticle, where the length was measured along the longest direction of each MRO. **c,** Partial PDFs of the fcc-, hcp-, bcc- and sc-like solute centres in the metallic glass nanoparticle, where the

maximum peak positions are located at 4.62, 4.77, 4.82 and 3.88 Å, respectively. Compared with the other three partial PDFs, the partial PDF of the sc-like solute centres (the purple curve) shows two peaks with the ratio of the second to the first peak position about $\sqrt{2}$. **d**, Distribution of sharing one, two, three, four and five atoms between neighbouring solute-centred clusters for the four types of the MRO.

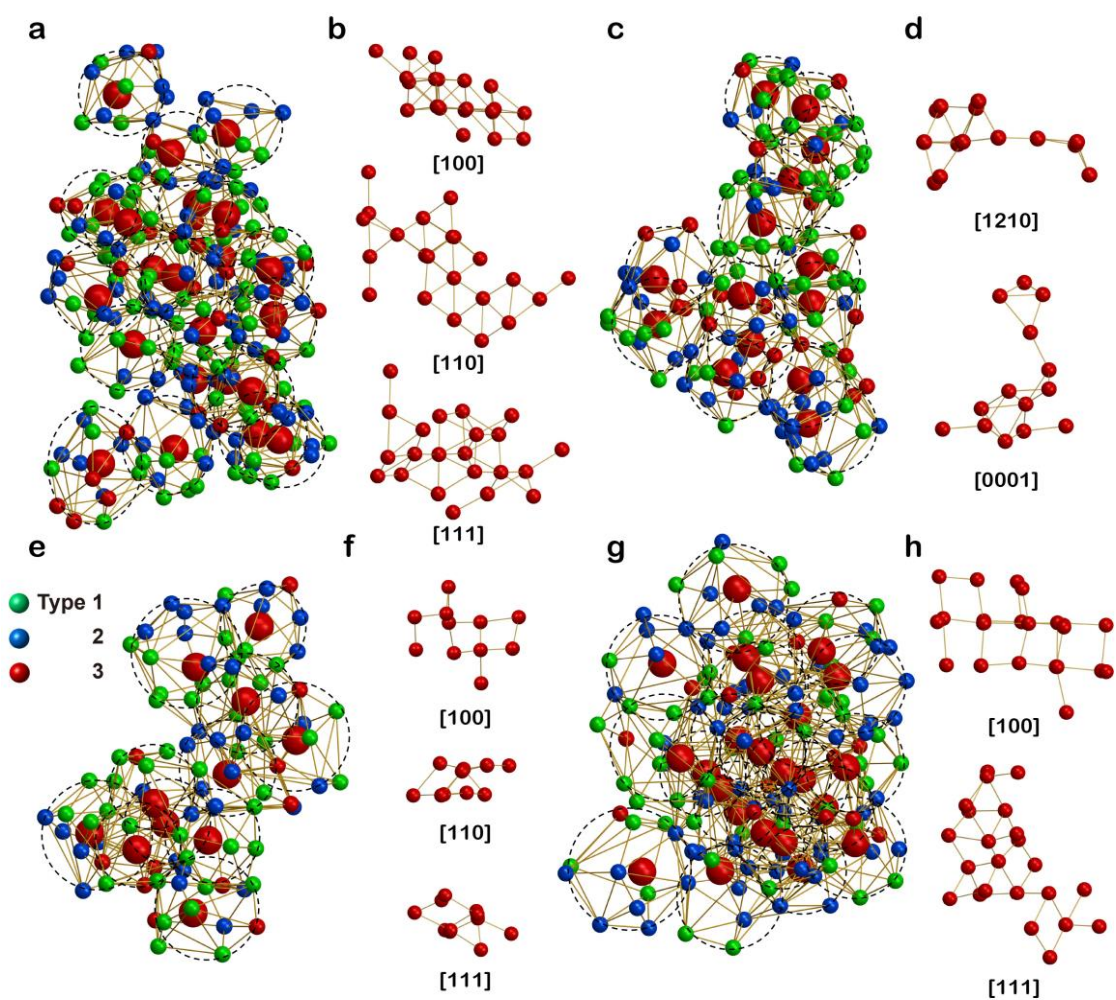

**Figure 5 | 3D atomic packing of four representative MRO.** Representative fcc- (**a**), hcp- (**c**), bcc- (**e**) and sc-like (**g**) MRO, consisting of 22, 14, 11 and 23 solute centres (large red spheres), respectively, where the solute-centred clusters (dashed circles) show only translational but no orientational order. To better visualize the crystal-like MRO, the solvent atoms are removed and the solute centres are orientated along the fcc (**b**), hcp (**d**),

bcc (**f**) and sc (**h**) zone axes, showing that the 3D shapes of the MRO are anisotropic and the networks significantly deviate from perfect crystal lattices.

## METHODS

**Sample preparation.** The multi-component metallic nanoparticle samples were synthesized using the thermal shock procedures published elsewhere[43]. Individual metal salts (chlorides or their hydrate forms) were dissolved in ethanol at a concentration of 0.05 mol/L. After complete dissolving with hydrochloric acid, the individual salt precursor solutions with different cations were mixed and sonicated for 30 minutes. The homogenously mixed precursor solution was loaded onto the carbon substrates[51] (reduced graphene oxide) and heated to a temperature as high as 1,763 K for 55 milliseconds (Supplementary Fig. 1a). The sample was suspended on a trench and connected with copper electrodes by silver paste for both heating and effective cooling as a giant heat sink. The thermal shock synthesis was triggered by electric Joule heating in an argon-filled glovebox using a Keithley 2425 SourceMeter where the high temperature and duration can be effectively controlled by tuning the input power and duration. The temperature of this process was monitored by a high-speed Phantom Miro M110 camera with a pixel size of 25 μm (Supplementary Video 1). The cooling rate was estimated to be ~5.1-6.9×$10^4$ K/s (Supplementary Fig. 1a), which, according to previous studies, can make metallic glasses[52,53]. The resulting nanoparticles on reduced graphene oxide were dispersed in ethanol with sonication. After deposited on to 5-nm-thick silicon nitride membranes, the nanoparticles were baked at 100 °C for 12 hours in vacuum to eliminate any hydrocarbon contamination. Both energy-dispersive X-ray and electron energy loss spectroscopy data show that the nanoparticles were still in metallic form and were not oxidized during the experiment (Supplementary Fig. 1b-q).

**Data acquisition.** A set of tomographic tilt series were acquired from seven nanoparticles using the TEAM 0.5 microscope with the TEAM stage[54]. Images were collected at 200 kV in ADF-STEM mode (Supplementary Table 1). To minimize sample drift, four sequential images per tilt angle were measured with a dwell time of 3 μs. To monitor any potential damage induced by the electron beam, we took 0° images before, during and after the acquisition of each tilt series and ensured that no noticeable structural change was observed for the seven nanoparticles. The total electron dose of each tilt series was estimated to be between 7×$10^5$ e$^-$/Å$^2$ and 9.5×$10^5$ e$^-$/Å$^2$ (Supplementary Table 1).

**Image pre-processing and denoising.** For each experimental tilt series, we performed the following procedure for image post-processing and denoising.

i) Image registration. At each tilt angle, we used the first image as a reference and calculated normalized cross-correlation between the reference and the other three images using a step size of 0.1 pixel[55]. These four images were aligned and summed to form an experimental image at that tilt angle.

ii) Scan distortion correction[38]. Two steps were used to correct the scan distortion for the experimental images. First, a set of low-magnification images were taken from nanoparticles and their positions were fitted with a Gaussian. Based on the geometric relation of the nanoparticles at different angles, the scan coil directions were calibrated to be perpendicular and equal in strength. Second, six high-magnification images were taken from a multi-component metallic nanoparticle and scan distortion parameters were estimated by minimizing the mean squared error of the common line of the six images. These scan distortion parameters were applied to the experimental images.

iii) Image denoising. The experimental images contain mixed Poisson and Gaussian noise and were denoised by the block-matching and 3D filtering (BM3D) algorithm[56], which has been demonstrated to be effective in reducing noise in AET[38,40,42]. The BM3D denoising parameters were optimized by the following three steps. First, Poisson and Gaussian noise level were estimated from the experimental tilt series. Second, several images were simulated based on a model nanoparticle, which has a similar size and elemental distribution as those of an experimental image. The same level of Poisson and Gaussian noise was added to the simulated images. Third, these noisy images were denoised by BM3D with different parameters. The denoising parameters corresponding to the largest cross-correlation coefficient between the denoised and the original images were chosen and applied to denoise the experimental images.

iv) Background subtraction and alignment. After denoising, a 2D mask was defined from each experimental image, which is slightly larger than the size of the nanoparticle. The background inside the mask was estimated by the discrete Laplacian in Matlab. After background subtraction, the experimental images of each tilt series were projected onto the tilt axis to produce a set of 1D curves (termed common lines). The images were aligned along the tilt axis by maximizing the cross-correlation between the common lines. The alignment of the images perpendicular to the tilt axis was achieved by the centre of mass method. The centres of mass of the images were calculated and the images were shifted so that all the centres of mass coincide with the origin. This image alignment method has been successfully used to achieve sub-pixel accuracy[34,36,40-42]. The Matlab data of the raw, processed and aligned experimental images are provided in Supplementary Information.

**The REal Space Iterative REconstruction (RESIRE) algorithm.** After post-processing and denoising, the experimental images were reconstructed by the RESIRE algorithm. The algorithm iteratively minimizes an error function defined by,

$$\varepsilon_\theta(O) = \frac{1}{2}\sum_{x,y}|\Pi_\theta(O)\{x,y\} - b_\theta\{x,y\}|^2 \qquad (1)$$

where $\varepsilon_\theta(O)$ is an error function of a 3D object ($O$) at tilt angle $\theta$, $\Pi_\theta(O)$ projects $O$ to generate a 2D image at angle $\theta$, $b_\theta$ is the experimental image at angle $\theta$, and $\{x,y\}$ is the coordinates. The minimization is solved via the gradient descent,

$$\nabla\varepsilon_\theta(O)\{u,v,w\} = \Pi_\theta(O)\{x,y\} - b_\theta\{x,y\} \quad \text{where } \begin{bmatrix}u\\v\\w\end{bmatrix} = R_\theta \begin{bmatrix}x\\y\\z\end{bmatrix} \text{ for some } z \quad (2)$$

where $\nabla$ represents the gradient and $R_\theta$ is the rotation matrix at tilt angle $\theta$, which transforms coordinates $\{x,y,z\}$ to $\{u,v,w\}$. The $j^{th}$ iteration of the RESIRE algorithm consists of the following four steps.

i) A set of images are calculated from the 3D object of the $j^{th}$ iteration using a Fourier method. The 3D object is first padded with zeros by properly choosing an oversampling ratio[57]. Applying the fast Fourier transform to the zero-padded object generates a 3D array in reciprocal space, from which a series of 2D

Fourier slices are obtained at different tilt angles. These 2D Fourier slices are inverted to a set of images via the inverse Fourier transform.

ii) The error function defined in equation 1 is calculated between the computed and experimental images.

iii) The gradient of the error function is computed for every voxel using equation 2.

iv) The 3D object of the (j+1)$^{th}$ iteration is updated by,

$$O^{j+1} = O^j - \frac{\Delta}{nN} \sum_\theta \nabla \varepsilon_\theta(O^j) \qquad (3)$$

where Δ is the step size (Δ = 2 was chosen for the reconstruction of our experimental data), *n* is the number of images and *N* is the dimension of each image (N×N). $O^{j+1}\{u, v, w\}$ is used as an input for the (j+1)$^{th}$ iteration.

The convergence of the algorithm is monitored by the R-factor,

$$R = \frac{1}{n} \sum_\theta \frac{\sum_{x,y} |\Pi_\theta(O)\{x,y\} - b_\theta\{x,y\}|}{\sum_{x,y} |b_\theta\{x,y\}|} \quad . \quad (4)$$

Usually, after several hundreds of iterations, the algorithm converges to a high-quality 3D reconstruction from a limited number of images. Both our numerical simulation and experimental results have indicated that RESIRE outperforms other iterative tomographic algorithms such as generalized Fourier iterative reconstruction[58] and simultaneous iterative reconstruction technique[59]. By avoiding iterating between real and reciprocal space, RESIRE can be applied to general sample geometry such as thin films and extended objects. The details of the RESIRE algorithm will be reported in a follow-up paper.

For each aligned experimental tilt series, we first ran RESIRE for 200 iterations. From the initial 3D reconstruction, we performed the angular refinement and spatial alignment for the experimental images[40,58]. For each experimental image, we determined the corresponding three Euler angles of the 3D reconstruction. We sequentially scanned each of the three Euler angles with a small angular increment. At each scanning step, we projected the 3D reconstruction back to calculate an image. The experimental image was shifted along the x and y-axis and aligned with the calculated one. An error metric, defined as the difference between the calculated and experimental image, was computed. After scanning all the three Euler angles, three optimal Euler angles was found with the smallest error metric. This procedure was iterated for all the experimental images until there was not further improvement, producing a set of spatially aligned experimental images and refined tilt angles. Next, the background of each experimental image was re-evaluated and re-subtracted. Using these experimental images with the refined tilt angles (Supplementary Fig. 4a), we ran another 200 iterations of RESIRE to obtain the final 3D reconstruction of each experimental tilt series (Supplementary Table 1). The source codes of RESIRE are provided in Supplementary Information.

**Determination of 3D atomic coordinates and species.** From each final 3D reconstruction, the atomic coordinates and species were identified using the following procedure[40,42].

i) Each 3D reconstruction was upsampled by a factor of 3 using the spline interpolation, from which all the local maxima were identified. Starting from the highest intensity peak, polynomial fitting[60] was performed on a 0.8×0.8×0.8 Å$^3$ (7×7×7 voxel) volume around each local maximum to locate the peak position. If the distance between the fitted peak position and existing potential atom positions is larger than or equal to 2

Å, it was listed as a potential atom. After repeating this step for all the local maxima, a list of potential atom positions was obtained. This method to trace the positions of potential atoms has previously been rigorously tested by using two independent experimental tilt series acquired from the same sample[42].

ii) A 3D difference map was generated by taking the difference between the 3D reconstruction and the list of the potential atoms. Based on the difference map, we manually adjusted a very small fraction of the atoms (167 out of 18356), which has been routinely used in protein crystallography[61].

iii) A K-mean clustering method[40,42,62] was used to classify three types of atoms and non-atoms (Co and Ni as type 1, Ru, Rh, Pd and Ag as type 2, and Ir and Pt as type 3) based on the integrated intensity of a 0.8 Å × 0.8 Å × 0.8 Å volume around each potential atom position. An initial atomic model with 3D atomic coordinates was determined from each 3D reconstruction.

iv) Due to the missing wedge problem and noise in the experimental images, there is local intensity variation in each 3D reconstruction. A local reclassification was iteratively performed to refine the type 1, 2 and 3 atoms. Each atom was defined as the centre of a 10-Å-radius sphere. The average intensity distribution of type 1, 2 and 3 atoms was computed within the sphere. The $L_2$ norm of the intensity distribution between the centre atom and the average type 1, 2 and 3 atom was calculated. The centre atom was assigned to the type with the smallest $L_2$ norm. The procedure was iteratively repeated until there were no further changes. The source codes for 3D atom tracing and classification are provided in Supplementary Information.

**Refinement of 3D atomic coordinates.** The 3D atomic coordinates were refined by minimizing the error between the calculated and measured images using the gradient descent[38,40,42]. Each atom was first fit with a 3D Gaussian function with a height $H$ and a width $B'$, where $H$ and $B'$ were considered the same for the same type of atoms. A 3D atomic model was obtained by,

$$O\{x,y,z\} = \sum_i H_i \exp\left[-\frac{|x-x_i|^2 + |y-y_i|^2 + |z-z_i|^2}{B'_i}\right] \quad (5)$$

where $x_i$, $y_i$, $z_i$, $H_i$ and $B'_i$ are the coordinates, height and standard deviation of the $i^{th}$ atom, respectively, $|x-x_i|, |y-y_i|, |z-z_i| \leq \rho$, and $\rho$ is a cut-off size of the 3D Gaussian function. From the 3D atomic model, a set of projection images were computed at different tilt angle $\theta$ by,

$$\Pi_\theta(O)\{u,v\} = \sum_w \sum_i H_i \exp\left[-\frac{|u-u_i|^2 + |v-v_i|^2 + |w-w_i|^2}{B'_i}\right] \quad (6)$$

$$\text{where } \begin{bmatrix} u_i \\ v_i \\ w_i \end{bmatrix} = R_\theta \begin{bmatrix} x_i \\ y_i \\ z_i \end{bmatrix} \quad \text{and} \quad |u-u_i|, |v-v_i|, |w-w_i| \leq \rho.$$

Substituting equation (6) into (1), an error function was calculated, from which the gradient descent method was used to search for the optimal atomic position at the (j+1)$^{th}$ iteration,

$$\{x_i, y_i, z_i\}^{j+1} = \{x_i, y_i, z_i\}^j - \Delta \sum_\theta [\Pi_\theta(O)\{u,v\} - b_\theta\{u,v\}] \nabla_i [\Pi_\theta(O)\{u,v\}] \quad (7)$$

Where $\nabla_i$ is the spatial gradient operator with respect to the atomic position $(x_i, y_i, z_i)$. The iterative refinement process was terminated when the $L_2$ norm error could not be further reduced.

**The local bond orientational order (BOO) parameters**. The local BOO parameters ($Q_4$ and $Q_6$) were calculated from the 3D atomic model of each nanoparticle using a method described elsewhere[63,64]. The $Q_4$ and $Q_6$ order parameters were computed up to the second shell with a shell radius set by the first valley in the RDF curve of the 3D atomic model. Figure 1f and Supplementary Fig. 2h-n show the distribution of the local BOO parameters of all the atoms in particles 1-7. To separate the amorphous structure from the crystal nuclei, we calculated the normalized local BOO parameter, defined as $\sqrt{Q_4^2 + Q_6^2}/\sqrt{Q_{4\,fcc}^2 + Q_{6\,fcc}^2}$, where $Q_{4\,fcc}$ and $Q_{6\,fcc}$ are the $Q_4$ and $Q_6$ value for a perfect fcc lattice. The normalized BOO parameter is between 0 and 1, where 0 means $Q_4 = Q_6 = 0$ and 1 represents a perfect fcc crystal structure. Based on the BOO parameters of a $Cu_{65}Zr_{35}$ metallic glass structure obtained from molecular dynamics simulations[65] (Supplementary Fig. 2o), we chose the normalized BOO parameter = 0.5 as a cut-off to separate crystal nuclei from amorphous structure (red curves in Fig. 1f and Supplementary Fig. 2h-n).

**3D precision estimation with multislice simulations**. A tilt series of 55 STEM images were calculated from the experimental 3D atomic model by using a fast multislice simulation software based on graphics processing unit[66]. At each refined experimental angle (Supplementary Fig. 4a), the experimental 3D atomic model was placed in a cuboidal super cell and the super cell was divided into multiple 2-Å-thick slices along the z-axis. The experiment parameters shown in Supplementary Table 1 (particle 1) were used for the multislice simulations. After using the parallel computing to perform the multislice simulations for all the angles, we calculated 55 multislice STEM images, each with 289×289 pixels and a pixel size of 0.347 Å. To account for the electron probe size and other incoherent effects, each multislice STEM image was convolved with a Gaussian kernel. Supplementary Fig. 4c and d show a representative experimental and multislice STEM image, respectively. An average R-factor between the 55 experimental and multislice images (defined in equation 4) was computed to be 14.96%, which, according to the crystallography standard[61], represents a good agreement between the two sets of images.

From the 55 multislice STEM images with angular errors (Supplementary Fig. 4a), we performed the 3D reconstruction and angular refinement with RESIRE (Supplementary Fig. 4b). After applying the atomic tracing, classification and refinement procedure to the reconstruction, we obtained a new 3D atomic model of the sample, consisting of 8438, 6905 and 3138 type 1, 2, and 3 atoms, respectively. We identified 7898, 6837, 3138 common pairs of type 1, 2 and 3 atoms, respectively, between the experimental and multislice atomic models based on the criterion of each common pair within a radius of 1.5 Å. The total common pairs of the three types of atoms are 17873, indicating that 97.37% of all atoms have been correctly identified. Supplementary Fig. 4d shows the distribution of the atomic deviation between all the common pairs with a root-mean-square deviation (i.e. 3D precision) of 21 pm.

**The radial distribution function (RDF) and partial pair distribution function (PDF).** The RDF was calculated for the 3D atomic model of each nanoparticle using the following procedure. i) The distance of all atom pairs in each 3D atomic model was computed and binned into a histogram. ii) The number of atom pairs in each bin was normalized with respect to the volume of the spherical shell corresponding to each bin. iii) The histogram was scaled so that the RDF approaches one for large separations. After plotting the RDF for each nanoparticle, the first valley of the RDF was used as the nearest neighbour cut-off distance to calculate the local BOO parameters (Fig. 1f and Supplementary Fig. 2h-n). By choosing the atoms in the metallic glass nanoparticle (particle 1) with the normalized BOO parameter < 0.5, we applied the above procedure to plot the RDF (Fig. 1g). For type 1, 2 and 3 atoms, we identified six sets of atoms pairs (type 11, 12, 13, 22, 23 and 33) in the nanoparticle. For each set of atom pairs, we used the above procedure to calculate the partial PDF shown in Fig. 1h.

**Voronoi tessellation and the coordination number (CN).** The analysis of Voronoi tessellation was performed by following the procedure published elsewhere[6], where the surface atoms of the nanoparticle were excluded. To reduce the effect of the experimental and reconstruction error on Voronoi tessellation, those surfaces with areas less than 1% of the total surface area of each Voronoi polyhedron were removed[9]. From the Voronoi tessellation, each polyhedron is designated by a Voronoi index $\langle n_1, n_2, n_3, n_4, \cdots \rangle$ with $n_i$ denoting the number of $i$-edged faces and the CN was calculated by $\sum_i n_i$.

**Quantification of the chemical SRO.** We used the Warren–Cowley parameters ($\alpha_{lm}$) to quantify the chemical SRO[67,68],

$$\alpha_{lm} = 1 - \frac{Z_{lm}}{\chi_m Z_l} \qquad (8)$$

where $l, m = 1, 2$ or $3$, $Z_{lm}$ is the partial CN of type $m$ atoms around type $l$ atoms, $\chi_m$ is the fraction of type $m$ atoms, and $Z_l$ is the total CN around type $l$ atoms. After excluding the surface atoms, we estimated $\chi_1$, $\chi_2$ and $\chi_3$ to be 42.97%, 38.28% and 18.75%, respectively. Using the partial CNs (Supplementary Fig. 5b), we calculated $\alpha_{11}$ = -0.11, $\alpha_{12}$ = 0.1, $\alpha_{13}$ = 0.05, $\alpha_{21}$ = 0.02, $\alpha_{22}$ = 0.01, $\alpha_{23}$ = -0.07, $\alpha_{31}$ = 0.03, $\alpha_{32}$ = -0.06, and $\alpha_{33}$ = 0.06, indicating that the type 11 and 23 bonds are favoured, but the type 12 and 33 bonds are unfavoured. These results are consistent with the observations that the type 23 bond is 0.06 Å shorter than the average type 2 and 3 bonds and the type 12 bond is 0.06 Å longer than the average type 1 and 2 bonds (Fig. 1h).

**Determination of the solute centres and the MRO.** A breadth-first-search algorithm[69,70] was implemented to search for the solute centres and the MRO using the following procedure. First, the algorithm identified the solute centres from type 3 atoms based on two criteria: i) each solute centre must fall within a 0.75 Å radius from an fcc, hcp, bcc or sc lattice point; and ii) each solute centre must have at least one neighbouring type 3 atom within the second-coordination-shell distance. Second, the identified solute centres were sorted out to generate a queue of the fcc-, hcp-, bcc- or sc-like MRO candidates. Third, starting from the largest MRO candidate (i.e. with the most solute centres), each candidate was classified as an MRO if it has at least five or more solute centres and none of the solute centres was already occupied by another MRO. If any solute centres were already occupied, they were removed from the MRO candidate and the candidate was refitted into the lattice vectors and added back into the queue. If two or more MRO candidates have the same number of solute centres, the one with the smallest error of fitting the solute centres into the lattice vectors was analysed first. This process was repeated until all the MRO was identified, where each solute centre can only belong to no more than one MRO. To corroborate our analysis, we repeated the above steps with a 1 Å and 0.5 Å radius cut-off and the corresponding MRO is shown in Supplementary Figs. 6 and 7, respectively.

An attempt was also made in searching for icosahedral-like MRO. The breadth-first-search algorithm[69,70] was used to find the MRO that falls within a 0.75 Å radius from the 12 vertices of an icosahedron. Because the icosahedron cannot be periodically packed in three dimensions, only the nearest neighbour vertices were searched, making the largest possible MRO have 13 solute centres (1 central solute centre plus 12 nearest neighbours). After performing the search, the resulting possible MRO has a mean value of 3.9, meaning on average each solute centre is connected to only 3 others when constrained to an icosahedron within the second coordination shell. Furthermore, although the largest possible MRO has 7 solute centres, none of these solute centres form 5-fold symmetry. We also repeated this analysis with a 1 Å radius cut-off. The mean value of solute centres becomes 4.5, the largest MRO has 8 solute centres, and there are 19 5-fold symmetries. The source codes for identifying the MRO are provided in Supplementary Information.

**Determination of the 3D atomic structure of an amorphous CuTa thin film**. The following procedure was used to experimentally resolve the 3D atomic positions in the CuTa thin film.

i) Sample preparation. CuTa thin films were fabricated *in-situ* in the sample chamber of the spin polarized low energy electron microscope (SPLEEM) at NCEM, where clean ultrahigh vacuum conditions remained in the low $10^{-9}$ torr range. Using thermal evaporation, CuTa thin films were deposited on $Si_3N_4$ substrates, which were maintained well below 150 K during sample fabrication. The growth rate of the thin films was in the range of 0.1 – 1 atomic monolayer per minute. After the fabrication of the CuTa thin films, a very thin carbon capping layer was deposited on the films to protect the samples from oxidation.

ii) Data acquisition. A tomographic tilt series was acquired from the CuTa thin film using TEAM I under ADF-STEM mode at 300 kV. To mitigate the sample drift, two images at each tilt angle were taken and then aligned to improve the signal-to-noise ratio. The tilt series consists of a total of 40 images with a tilt range from -67.9º to 64.9º (Supplementary Fig. 8). As the CuTa film is thinner than ~6 nm, 40 experimental images are sufficient to produce a good 3D reconstruction. The total electron dose of the dataset is $4.8 \times 10^5$ e/Å$^2$. All the experimental parameters of the tilt series can be found in Supplementary Table 1.

iii) Image alignment. All the image pre-processing and denoising steps for the analysis of the CuTa thin film are similar to those of the metallic glass nanoparticle, except for image alignment. We first used the cross-correlation between the neighbouring images to roughly align the CuTa images. Next, we searched for some reference markers, which can be either created by adding some small nanoparticles or based on features in the sample. In this experiment, we chose an isolated region in the images and aligned them using the centre of mass and common line method[34,36]. After obtaining the 3D reconstruction, we further refined the alignment by projecting the reconstruction back to generate images and comparing them with the experimental ones. This process was repeated until no further improvement could be made.

iv) 3D reconstruction, atomic tracing and refinement. Using RESIRE, we first performed a large volume reconstruction of the CuTa thin film from the aligned images. Based on the thickness variation of the thin film, we applied scanning AET[41] to conduct multiple local volume reconstructions and then patched them together to produce a full 3D reconstruction. Scanning AET has been previously demonstrated to be effective in improving the 3D reconstruction of 2D materials and/or thin film samples[41]. From the full 3D reconstruction, we projected it back to generate images and use them to perform the angular refinement and spatial alignment. We iteratively repeated the process until there were no further changes. After obtaining the final 3D reconstruction, we traced the Cu and Ta atoms based on the integrated intensity difference between the two types of atoms. The 3D atomic positions were refined to produce a final 3D atomic model of the CuTa thin film (Supplementary Fig. 9).

**Supplementary Table and Figures**
**Supplementary Table 1 | AET data collection, processing, reconstruction, refinement and statistics.**

|  | Particle 1 | Particle 2 | Particle 3 | Particle 4 | Particle 5 | Particle 6 | Particle 7 | CuTa film |
|---|---|---|---|---|---|---|---|---|
| **Data collection and processing** | | | | | | | | |
| Voltage (kV) | 200 | 200 | 200 | 200 | 200 | 200 | 200 | 300 |
| Convergence semi-angle (mrad) | 25 | 25 | 25 | 25 | 25 | 25 | 25 | 17.1 |
| Probe size (Å) | 0.8 | 0.8 | 0.8 | 0.8 | 0.8 | 0.8 | 0.8 | 0.7 |
| Detector inner angle (mrad) | 38 | 38 | 38 | 38 | 38 | 38 | 38 | 30 |
| Detector outer angle (mrad) | 190 | 190 | 190 | 190 | 190 | 190 | 190 | 195 |
| Depth of focus (nm) | 8 | 8 | 8 | 8 | 8 | 8 | 8 | 14 |
| Pixel size (Å) | 0.347 | 0.347 | 0.347 | 0.347 | 0.347 | 0.347 | 0.347 | 0.322 |
| # of images | 55 | 51 | 54 | 54 | 53 | 53 | 55 | 40 |
| Tilt range (°) | -69.4 +72.6 | -69.3 +63.4 | -72.5 +63.4 | -71.7 +69.4 | -69.6 +74.0 | -71.7 +69.4 | -69.5 +72.0 | -67.9 +64.9 |
| Total electron dose ($10^5$ e/Å$^2$) | 9.5 | 7.1 | 7.6 | 7.6 | 7.4 | 7.4 | 7.7 | 4.8 |
| **Reconstruction** | | | | | | | | |
| Algorithm | RESIRE | RESIRE | RESIRE | RESIRE | RESIRE | RESIRE | RESIRE | RESIRE |
| Oversampling ratio | 4 | 4 | 4 | 4 | 4 | 4 | 4 | 3 |
| Number of iterations | 200 | 200 | 200 | 200 | 200 | 200 | 200 | 500 |
| **Refinement** | | | | | | | | |
| $R_1$ (%)[a] | 9.54 | 12.32 | 11.57 | 10.90 | 9.38 | 7.62 | 7.65 | 10.58 |
| R (%)[b] | 6.45 | 9.87 | 8.98 | 7.68 | 7.46 | 5.66 | 5.73 | 8.61 |
| B' factors (Å$^2$) | | | | | | | | |
|   Type 1 atoms | 48.9 | 55.4 | 48.4 | 44.0 | 35.5 | 55.6 | 35.5 | 45.2 |
|   Type 2 atoms | 46.4 | 47.3 | 44.3 | 44.0 | 34.5 | 58.0 | 34.5 | 45.2 |
|   Type 3 atoms | 52.3 | 37.4 | 39.6 | 40.9 | 34.6 | 42.1 | 34.6 | NA |
| **Statistics** | | | | | | | | |
| # of atoms | | | | | | | | |
|   Total | 18356 | 2063 | 3447 | 4661 | 7739 | 6648 | 6037 | 14582 |
|   Type 1 | 8322 | 648 | 1116 | 1079 | 2158 | 1579 | 1446 | 1808(Cu) |
|   Type 2 | 6896 | 937 | 1327 | 1906 | 2750 | 2430 | 2045 | 12774(Ta) |
|   Type 3 | 3138 | 478 | 1004 | 1676 | 2831 | 2639 | 2546 | |

[a]The $R_1$-factor is defined as equation 5 in ref. 40. [b]The R-factor is defined in equation 4 in Methods.

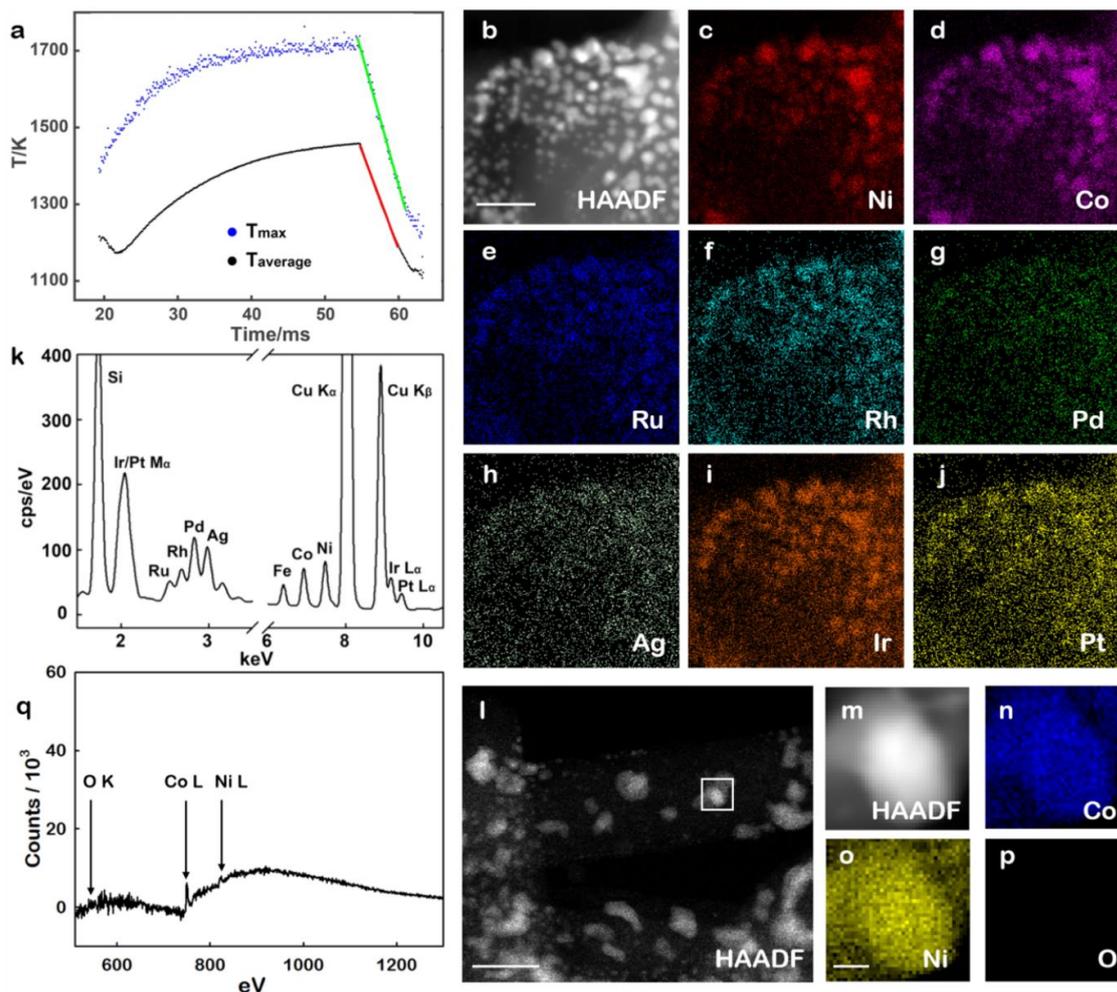

**Supplementary Fig. 1 | Cooling rate measurement, energy-dispersive X-ray (EDX) and electron energy loss spectroscopy (EELS) maps of the nanoparticles**. **a,** The cooling rate for the average and maximum temperature curves was measured to be 51000 K/s (the slope of the red line) and 69000 K/s (the slope of the green line), respectively. **b,** Low-resolution ADF-STEM image of the nanoparticles. EDX maps show the distribution of Ni (**c**), Co (**d**), Ru (**e**), Rh (**f**), Pd (**g**), Ag (**h**), Ir (**i**) and Pt (**j**). **k,** The EDX spectrum of all the elements shown in the images (**c-j**), where cps stands for counts per second. **l,** Low-resolution ADF-STEM image of a large area, in which the white square indicates the aggregation of several nanoparticles used for the EELS measurement. **m**, ADF-STEM image of the white square region. **n–p,** EELS maps show the distribution of Co (**n**), Ni (**o**) and O (**p**) in the region. **q**, EELS spectrum obtained from (**n–p**). No oxygen signal was detected in the EELS map or spectrum. Scale bars, 20 nm in (**b**), 100 nm in (**l**) and 10 nm in (**o**).

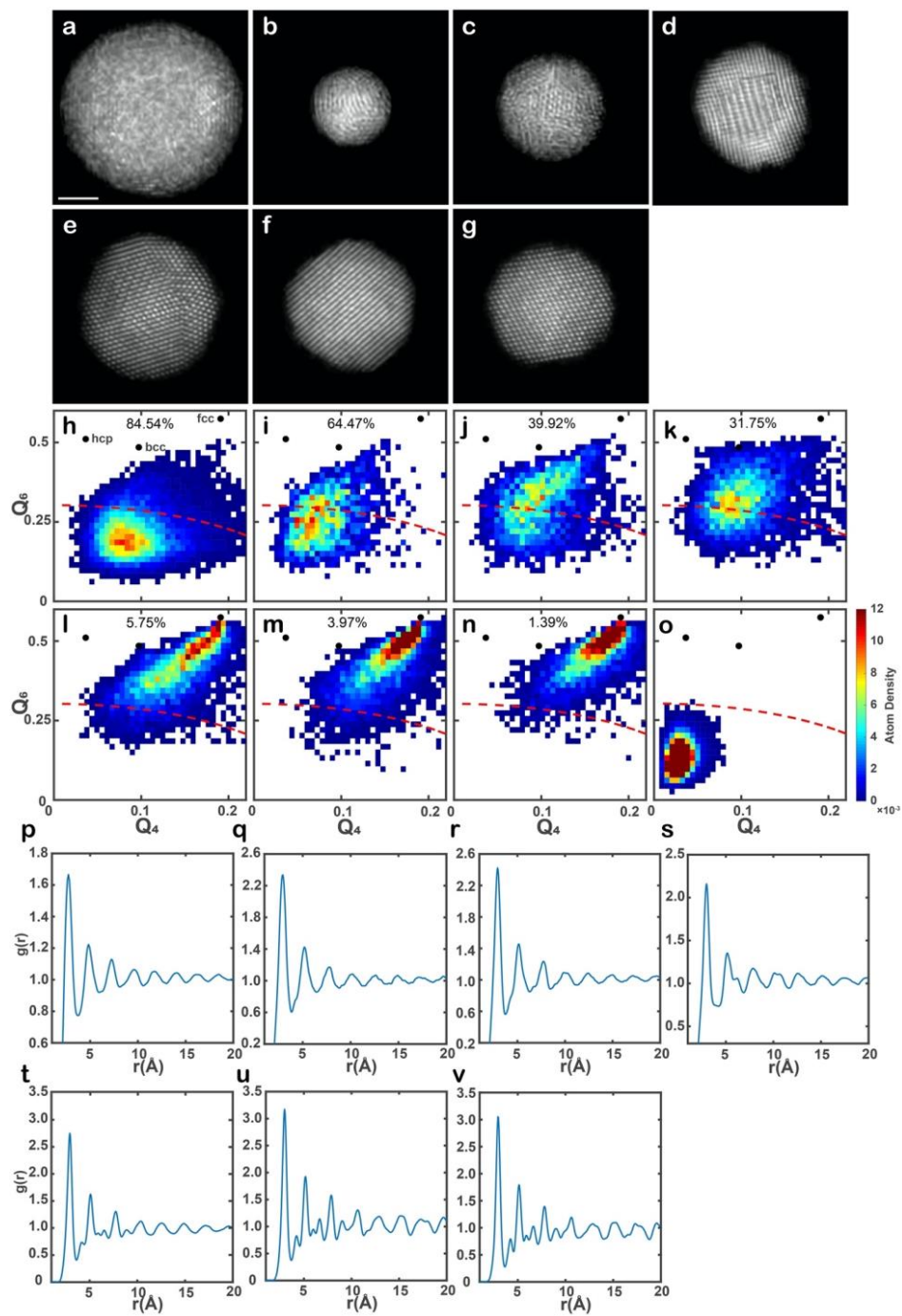

**Supplementary Fig. 2 | Analysis of seven multi-component metallic nanoparticles. a-g**, Representative ADF-STEM images of particles 1-7, respectively. Scale bar, 2 nm. **h-n**, Local BOO parameters of all the atoms in particles 1-7, where the dashed red lines correspond to the normalized BOO parameter = 0.5. The percentage on the top of each panel shows the fraction of disordered atoms in each particle. **o**, Local BOO parameters of a $Cu_{65}Zr_{35}$ metallic glass obtained from molecular dynamics simulations[65] as a reference, from which the normalized BOO parameter = 0.5 (dashed red curves) was chosen as a cut-off to separate crystal nuclei from amorphous structure. **p-v**, RDFs of all the atoms in particles 1-7, respectively. With the decrease of the fraction of disordered atoms in the nanoparticles, the peaks in the RDFs become narrower and new peaks corresponding to different crystal planes appear.

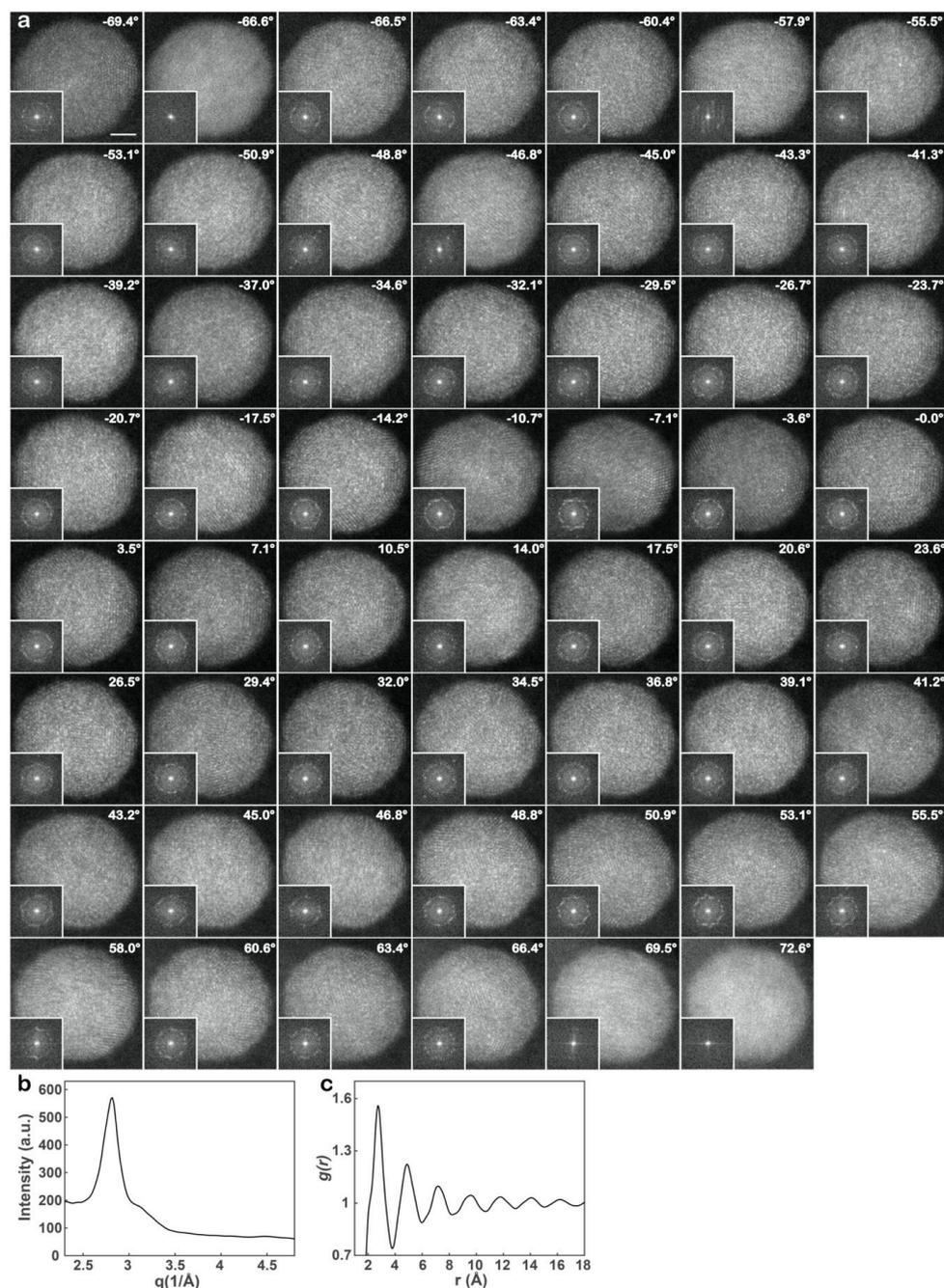

**Supplementary Fig. 3 | Experimental tomographic tilt series of the multi-component metallic glass nanoparticle (particle 1) with crystal nuclei. a**, 55 raw ADF-STEM images of the nanoparticle with a tilt range from -69.4° to +72.6°. The diffraction patterns of the images are shown in the insets, where the amorphous halo is visible. Some crystalline features are visible in several images and the diffraction patterns. Scale bar, 2 nm. **b**, Radial intensity distribution as a function of the spatial frequency ($q$), obtained from the average diffraction pattern (Fig. 1b). **c**, The RDF of the nanoparticle derived from the radial intensity distribution using the SUePDF software[71]. The first, second, third, fourth and fifth peak positions were determined to be 2.74, 4.74, 5.37, 7.12 and 9.6 Å, respectively, which agree well with those in Fig. 1g. The ratios of the second, third, fourth and fifth to the first peak position of the RDF are 1.73, 1.96, 2.60 and 3.50, respectively, which are consistent with those of metallic glasses[45,46].

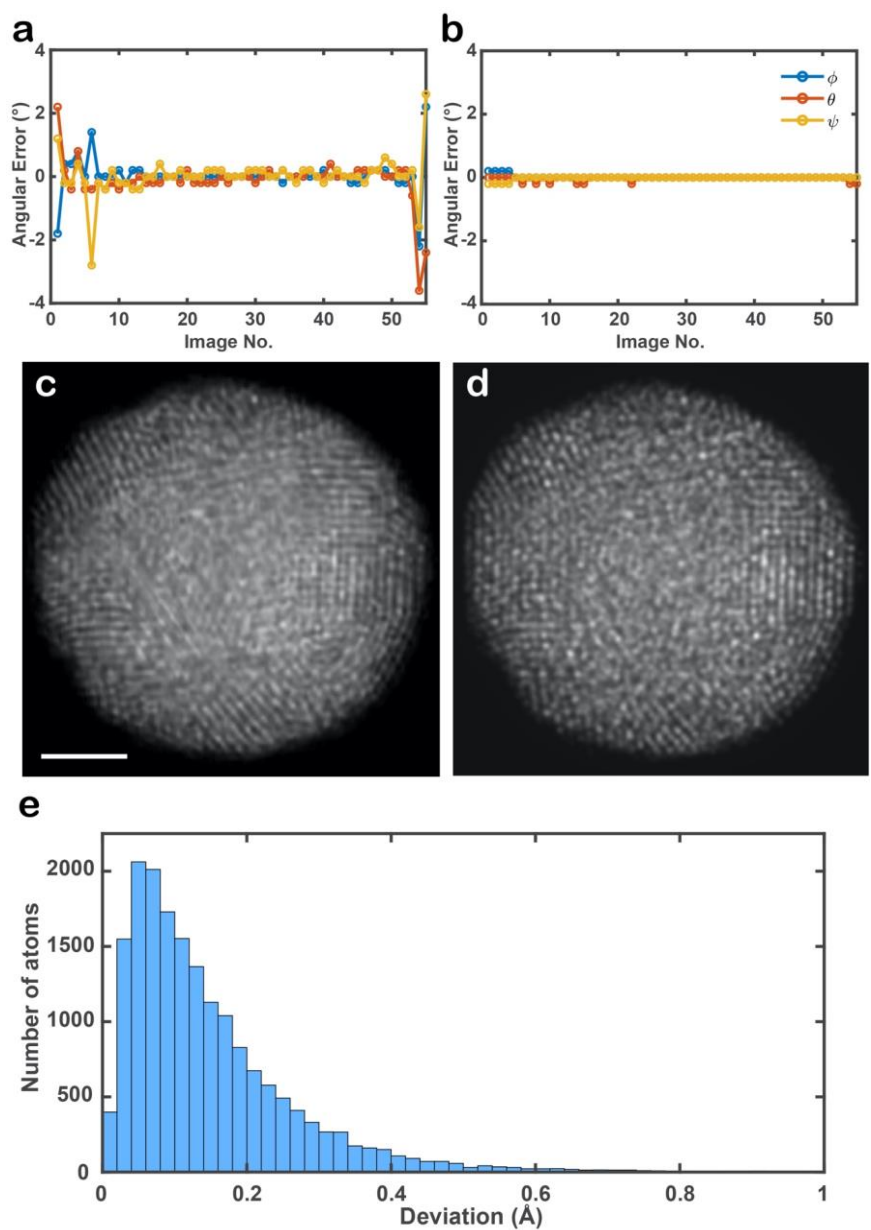

**Supplementary Fig. 4 | Angular errors in the experimental images and verification of the experimental 3D atomic model using multislice simulations. a**, Angular errors in the experimental images determined by the angular refinement procedure (Methods), where the colour dots and lines represent the deviation of the three Euler angles *(ϕ, θ and φ)* from the correct ones (0°) at each tilt angle. These angular errors were taken into account in the multislice simulations. **b**, The angular errors were correctly refined in the 3D reconstruction of the 55 multislice images using RESIRE (Methods). After the angular refinement, the largest error is only 0.2°. Comparison between a representative experimental (after denoising) (**c**) and a multislice image (**d**) at 0°. To account for the source size and incoherent effects, each multislice image was convolved with a Gaussian function (Methods). **e**, Histogram of the root mean square deviation of the atomic positions between the experimental atomic model and that obtained from 55 multislice images. Scale bar, 2 nm.

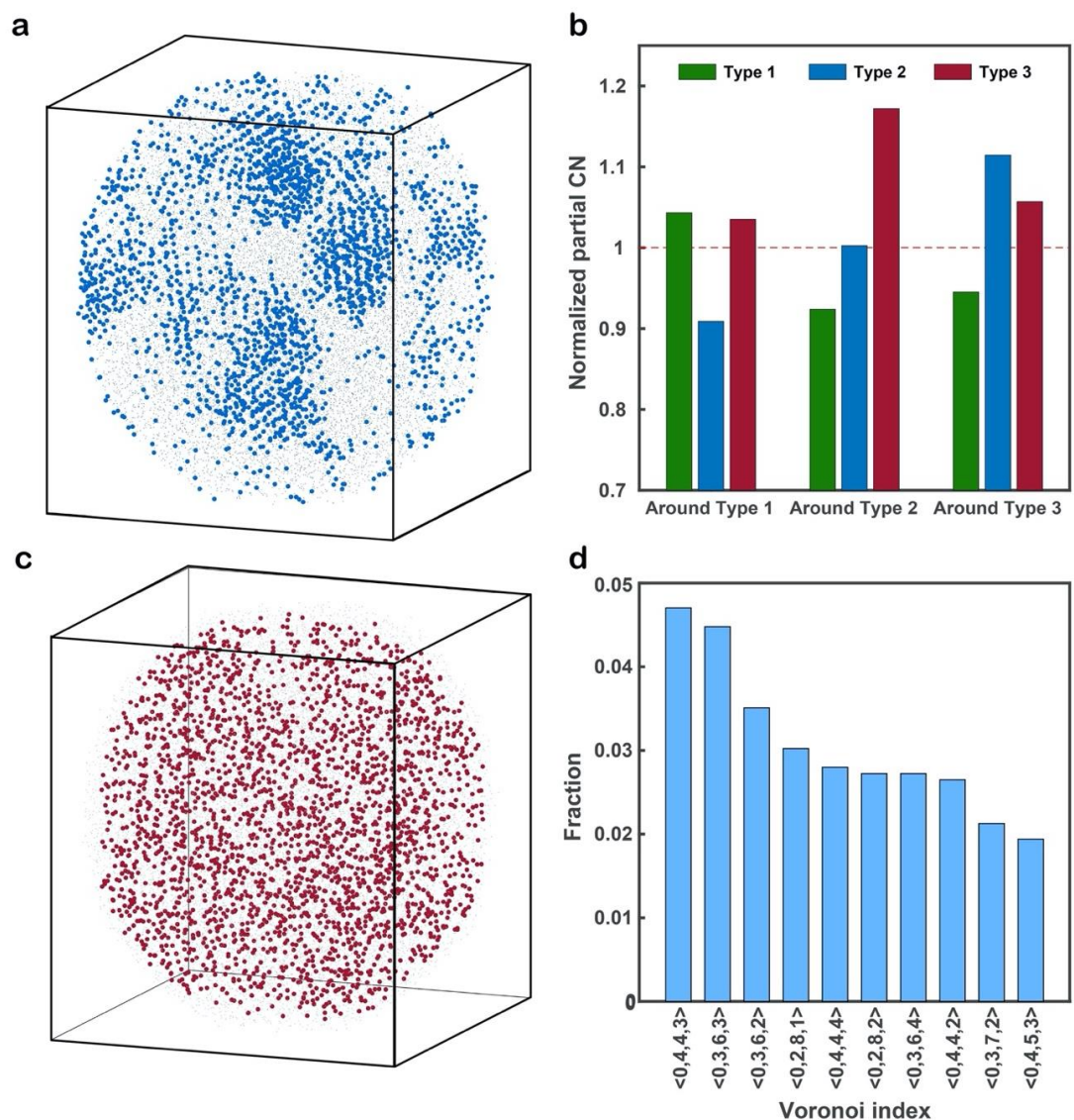

**Supplementary Fig. 5 | 3D distribution of the crystal nuclei in the nanoparticle, the partial CNs and the Voronoi polyhedra of the solute-centred clusters. a**, 3D distribution of the atoms with the normalized BOO parameter $\geq 0.5$, revealing 15.46% of the total atoms forming crystal nuclei in the nanoparticle. **b**, Normalized partial CNs around type 1, 2 and 3 atoms. **c**, 3D distribution of the 2682 solute centres (red dots), which are between the first (3.78 Å) and the second minimum (6.09 Å) of the RDF curve (Fig. 1g). **d**, Ten most abundant Voronoi polyhedra of the solute-centred clusters.

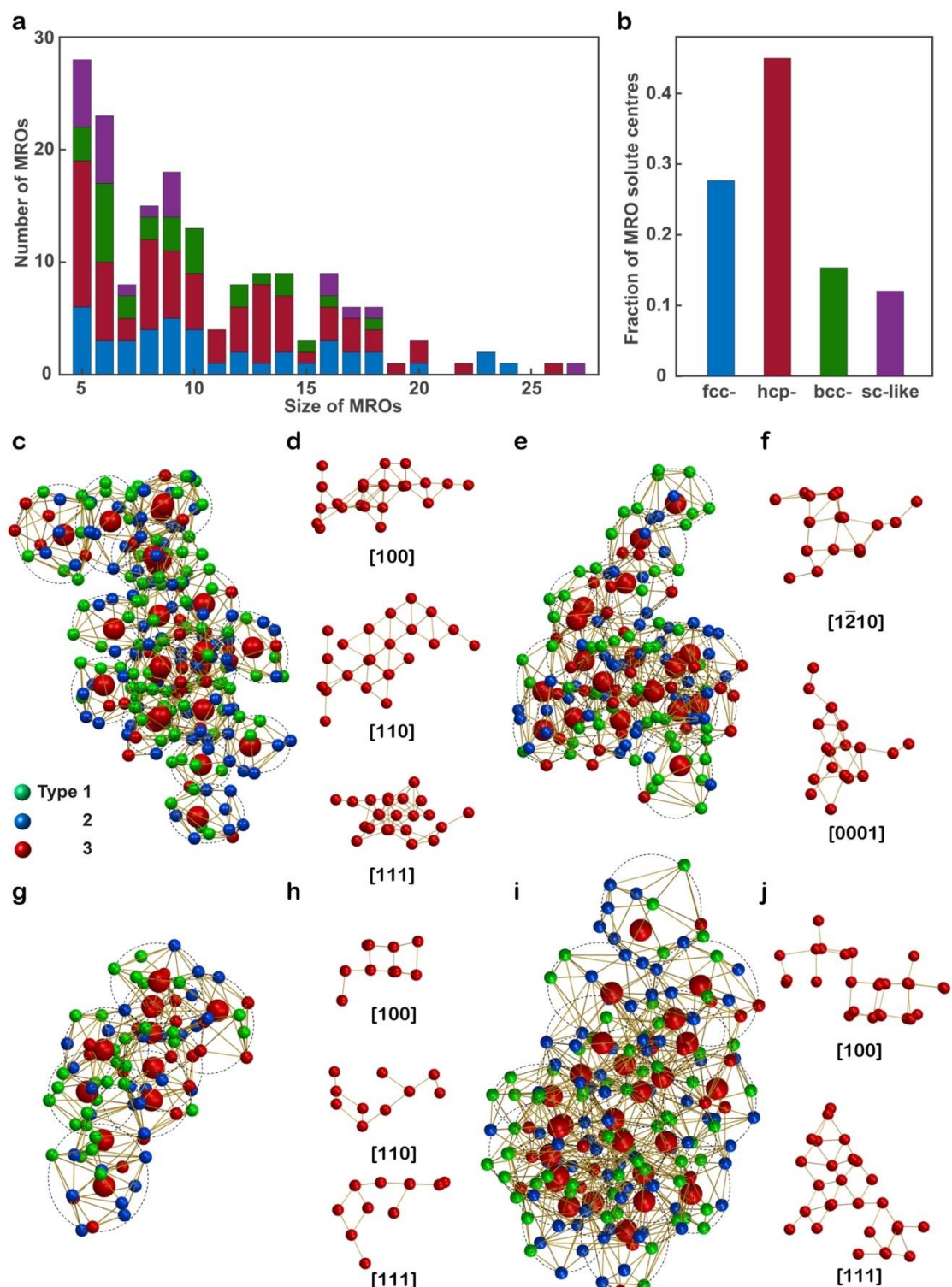

**Supplementary Fig. 6 | Identification of MROs with a 1 Å radius cut-off. a**, Histogram of the four types of MROs − fcc- (in blue), hcp- (in red), bcc- (in green) and sc-like (in purple) − as a function of the size (i.e. the number of solute centres). **b**, The fraction of the four MRO solute centre atoms. Representative fcc- (**c**), hcp- (**e**), bcc- (**g**) and sc-like (**i**) MROs, consisting of 23, 18, 10 and 27 solute centres (large red spheres), respectively. The solute centres are orientated along the fcc (**d**), hcp (**f**), bcc (**h**) and sc (**j**) zone axes.

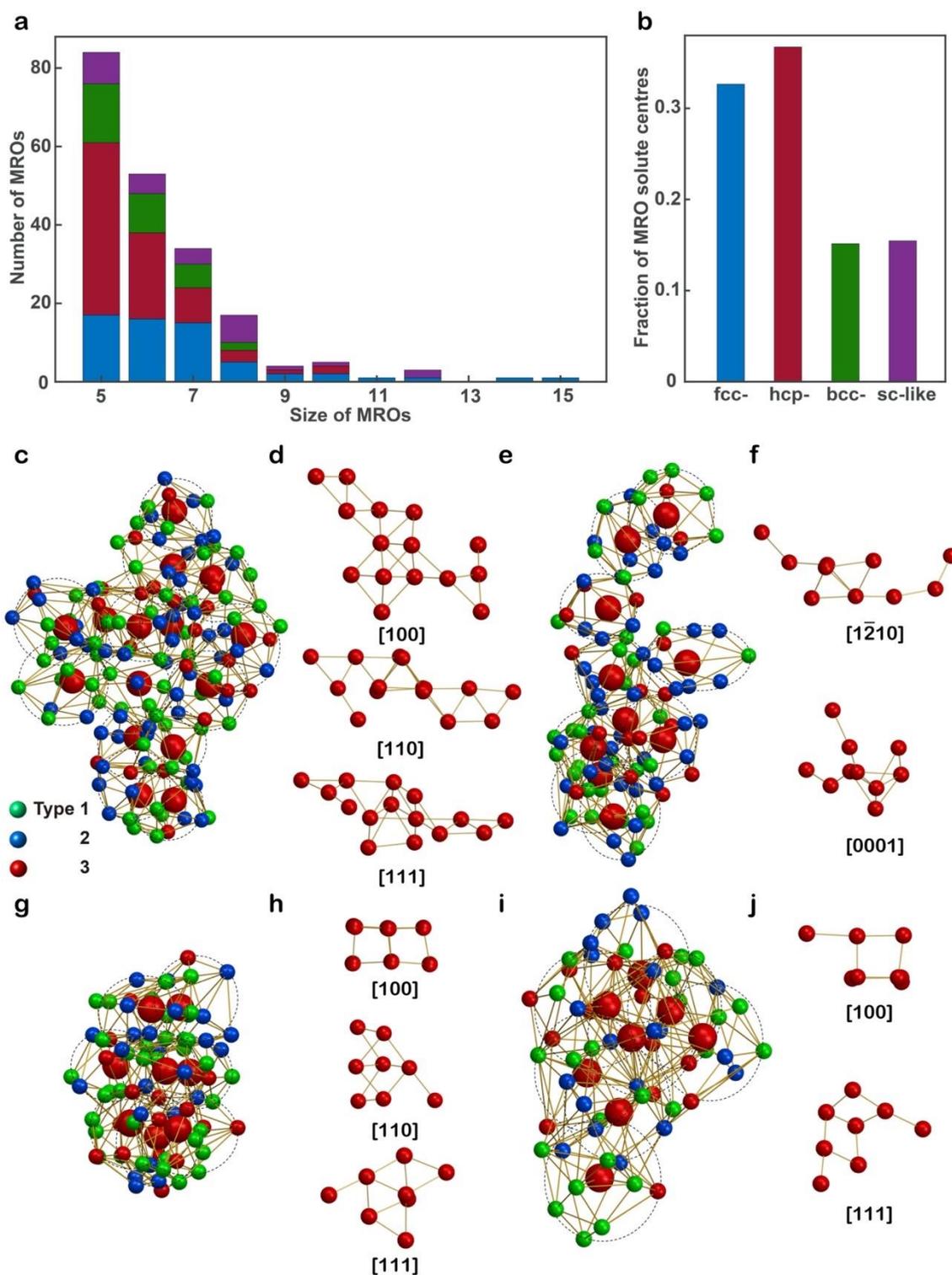

**Supplementary Fig. 7 | Identification of MROs with a 0.5 Å radius cut-off. a**, Histogram of the four types of MROs – fcc- (in blue), hcp- (in red), bcc- (in green) and sc-like (in purple) – as a function of the size. **b**, The fraction of the four MRO solute centre atoms. Representative fcc- (**c**), hcp- (**e**), bcc- (**g**) and sc-like (**i**) MROs, consisting of 15, 10, 8 and 8 solute centres (large red spheres), respectively. The solute centres are orientated along the fcc (**d**), hcp (**f**), bcc (**h**) and sc (**j**) zone axes.

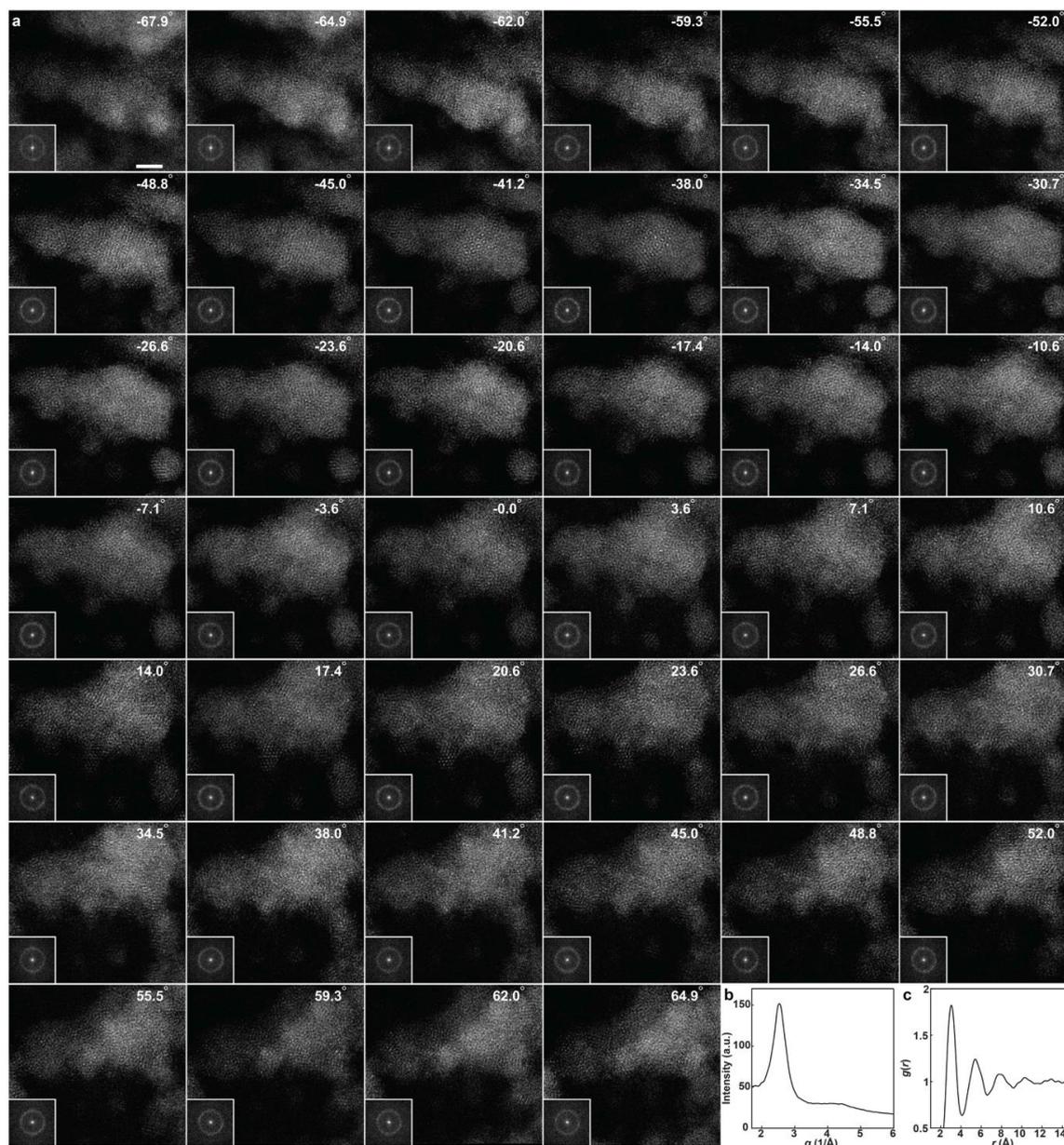

**Supplementary Fig. 8 | Tomographic tilt series of an amorphous CuTa thin film. a**, ADF-STEM images of a portion of the CuTa thin film. The insets show the diffraction patterns calculated from the experimental images, in which the diffuse halos are clearly visible. **b**, Radial intensity distribution as a function of the spatial frequency ($q$), derived by averaging all the diffraction patterns. **c**, The RDF derived from the radial intensity distribution using the SUePDF software[71]. The peak positions of the RDF agree with those in Supplementary Fig. 9g.

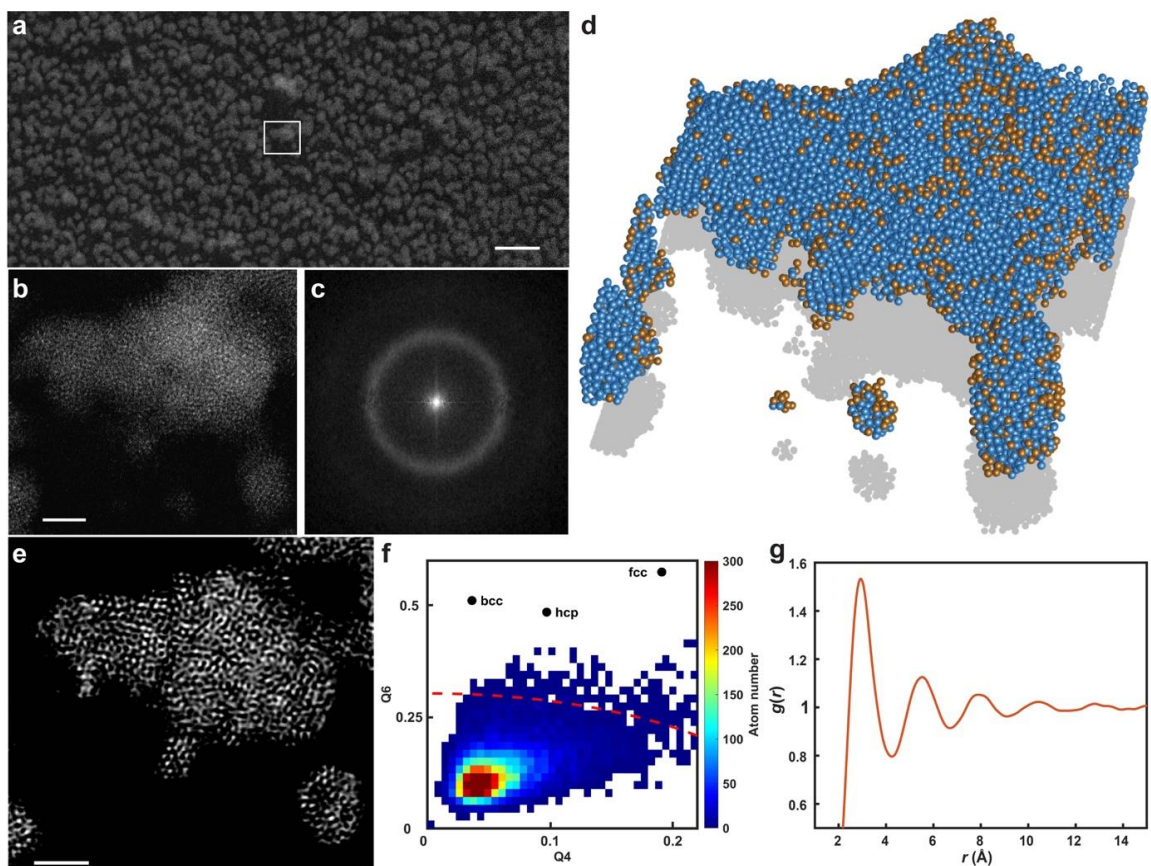

**Supplementary Fig. 9 | Determination of the 3D atomic structure of the amorphous CuTa thin film. a**, A large field of view of the amorphous CuTa. **b**, Magnified white square region in (**a**). **c**, Average diffraction pattern of all the experimental images. **d**, 3D atomic model of the portion of the CuTa thin film with a total of 1808 Cu (in gold) and 12774 Ta (in blue) atoms, determined from the tilt series shown in Supplementary Fig. 8a (Methods). As the CuTa film is thinner than ~6 nm, 40 experimental images are sufficient to produce a good 3D reconstruction. **e**, A 2-Å-thick internal slice of the 3D reconstruction of the amorphous CuTa thin film, showing the disordered atomic structure. **f**, Local BOO parameters of the 3D atomic model, where only 0.47% of the total atoms with the normalized BOO parameter ≥ 0.5 form crystal nuclei. **g**, RDF of the disordered atoms with the normalized BOO parameter < 0.5. The peak positions of the RDF agree with those shown in the inset of Supplementary Fig. 8c. Scale bars, 30 nm in (**a**), 2 nm in (**b**) and (**e**).